\documentclass[aps,prx,twocolumn,showpacs,superscriptaddress]{revtex4-2}  % for review and submission
\usepackage{graphicx}  % needed for figures
\usepackage{dcolumn}   % needed for some tables
\usepackage{bm}        % for math
\usepackage{amssymb}   % for math
\usepackage{amsmath}   % for math

\usepackage{verbatim}  % comments 
\usepackage{xcolor}    % color of text

\begin{document}
\title{Charge-conserving equilibration of quantum Hall edge states}

\author{Edvin G. Idrisov}
%\affiliation{Department of Physics and Materials Science, University of Luxembourg, L-1511 Luxembourg, Luxembourg}
%\affiliation{\mbox{}}
\affiliation{Department of Physics, United Arab Emirates University, P.O. Box 15551 Al-Ain, United Arab Emirates}

\author{Ivan P. Levkivskyi}
\affiliation{Hopper, Inc., Dublin, Ireland}

\author{Eugene V. Sukhorukov}
\affiliation{D\'epartement de Physique Th\'eorique, Universit\'e de Gen\`eve, CH-1211 Gen\`eve 4, Switzerland}
\date{\today}

% The following information is for internal review, please remove them for submission
%\widetext
%\leftline{Version xx as of \today}
%\leftline{Primary authors: Joe E. Physics}
%\leftline{To be submitted to (PRL, PRD-RC, PRD, PLB; choose one.)}
%\leftline{Comment to {\tt d0-run2eb-nnn@fnal.gov} by xxx, yyy}
%\centerline{\em D\O\ INTERNAL DOCUMENT -- NOT FOR PUBLIC DISTRIBUTION}

% the following line is for submission, including submission to the arXiv!!
%\hspace{5.2in} \mbox{Fermilab-Pub-04/xxx-E}

%\title{Template for PRL/PRD Papers}
%\input author_list.tex       % D0 authors (remove the first 3 lines
                             % of this file prior to submission, they
                             % contain a time stamp for the authorlist)
                             % (includes institutions and visitors)
%\date{\today}

\begin{abstract}
We address the experimentally relevant situation, where a non-equilibrium state is created by  injecting charge current into a chiral quantum Hall edge state. We show that the commonly accepted picture of the full equilibration of a non-equilibrium state at finite distances contradicts to the charge conservation requirement. We propose and solve  the transmission line model that accounts for the local equilibration process and the charge and energy conserving dynamics of the collective mode. We find that the correction of the electron distribution function to its eventual equilibrium form scales down slowly as $1/\sqrt{L}$ at long distances $L$.
\end{abstract}

\maketitle

%\section{\label{sec:level1}First-level heading}
% sections are not used for PRL papers

\section{Introduction}
\label{Introduction}
\begin{figure}
\includegraphics[width=\columnwidth]{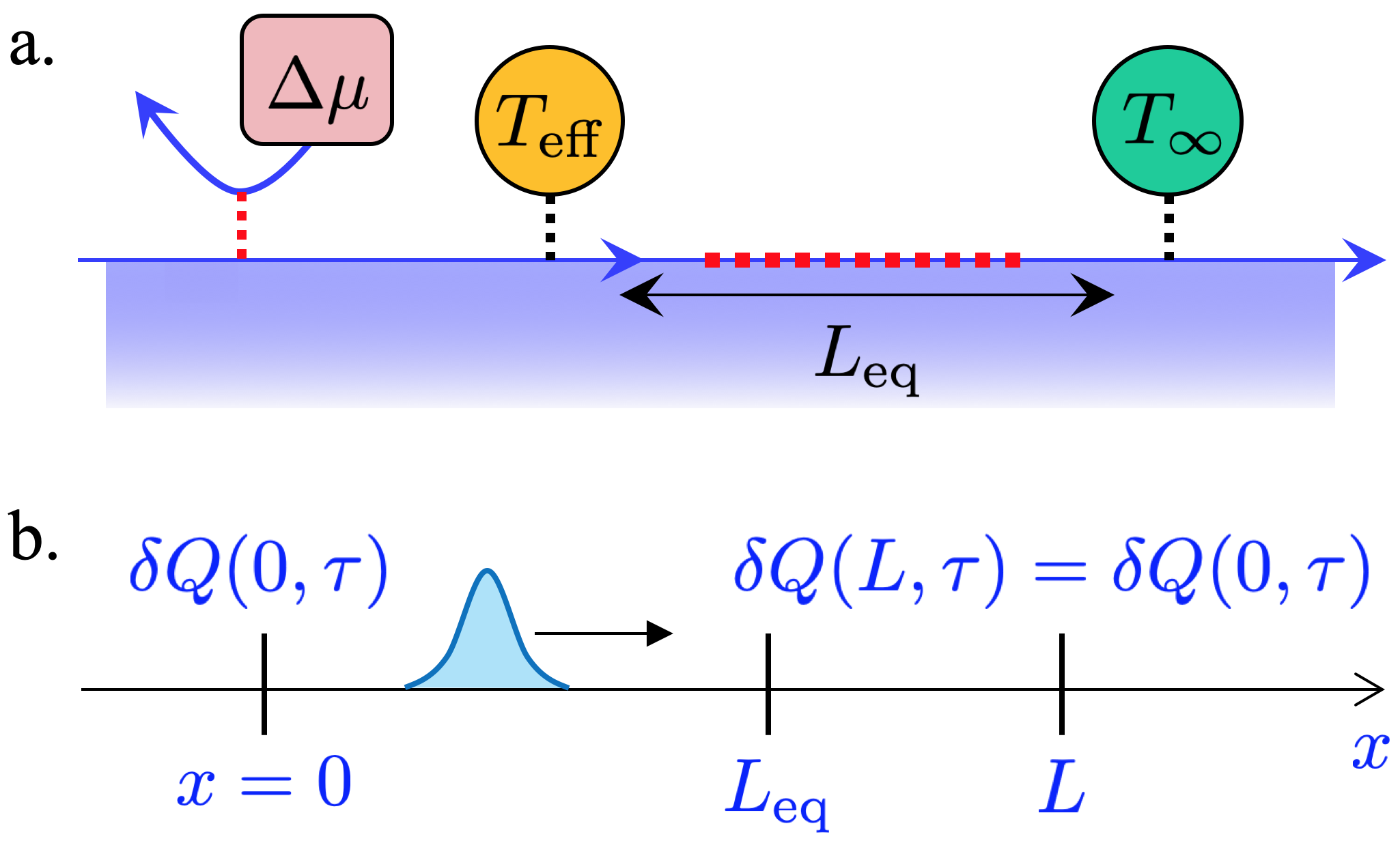}
\caption{\label{fig:one} (a) A  typical experimental setup for studying relaxation processes at QH edge state is schematically shown. A non-equilibrium state with the effective noise temperature $T_{\rm eff}$ is created by injecting a charge current, e.g., via a voltage biased QPC.  The state is detected downstream the injection point by measuring the electron distribution function or the current noise power. It is typically assumed that the state reaches local equilibrium with the temperature $T_\infty$ after characteristic distances $L_{\rm eq}$. (b) These schematics illustrate the argument presented in the main text that the intuitive picture of the equilibration process at the edge is incorrect, since it contradicts the charges conservation requirement.}
\end{figure}

The experimental progress in the field of hybrid mesoscale systems based on chiral quantum Hall (QH) edge states has triggered renewed interest in the phenomena of phase coherence \cite{Bisognin2019,Pierre1,Heiblum0,Nakamura2020}, charge and heat quantization \cite{Jezouin2016,IdrisovThermalDecay,ArturEquil,Sivre2018,Fujisawa1,Johan,Sivre2019}, relaxation and
equilibration \cite{Altimiras2010,Chalker,IvanRelaxation,Slobodeniuk1,Fujisawa2,Fujisawa3,Rodriguez2020,Rosenblatt} and entanglement \cite{Grenier,Jullien,Bauerle,Fletcher2019}, which are related to the fundamental problems of mesoscopic physics. In QH systems these phenomena are often observed by injecting electrons into a QH edge state using a quantum point contact (QPC) \cite{AltimirasRelaxation,Sueur}, quantum dot (QD) \cite{Roche1} or a mesoscopic Ohmic contact  (a metallic reservoir  of finite charge capacitance) \cite{Sivre2018,Sivre2019} (see Fig.\ \ref{fig:one}a) and detecting the charge current, heat current, current noise or an electron distribution function \cite{AltimirasRelaxation,Sueur} downstream of the injection  point. It is commonly assumed that after a relatively short distance from the injection point the state reaches its local equilibrium with the temperature that depends on the injected heat (in the case where the edge state is thermally isolated from the bulk of the system). Such a point of view seems to be supported by the studies of finite lifetime of excitations in one-dimensional systems \cite{Imambekov}.

However, the assumption of full equilibration of a one-dimensional state at distances longer than the  characteristic length $L_{\rm eq}$  leads to the following charge conservation paradox (see Fig.\ \ref{fig:one}b). Assuming the state is equilibrium at the distance $L>L_{\rm eq}$ from the injection point, the spectral density of current fluctuations $\delta j(L,t) = j -\langle j\rangle$ (noise power) at zero frequency satisfies the fluctuation-dissipation relation 
\begin{equation}
\label{FDT}
S(L,0)\!=\!\!\int\!\! dt\langle\delta j(L,t)\delta j(L,0)\rangle=G_qT_\infty,\;\;\; L>L_{\rm eq},
\end{equation}
where $T_\infty$ is the temperature of the final state, and $G_q=e^2/2\pi\hbar$ is the conductance quantum. On the other hand, the same quantity can be expressed in terms of the variance of the charge $\delta Q(L,\tau)=\int_0^\tau  dt\delta j(L,t)$ transmitted through the cross-section $x=L$ for a long time $\tau$. Indeed, 
\begin{equation}
\label{Variance of charge 1}
\frac{\langle [\delta Q(L,\tau)]^2\rangle}{\tau}=
\int d\omega\, S(L,\omega) \cdot  \frac{1-\cos(\omega \tau)}
{\pi \omega^2 \tau}.
\end{equation}
%\begin{align}
%\nonumber
%\frac{\langle [\delta Q(L,\tau)]^2\rangle}{\tau}&=\frac{1}
%{\tau} \int_0^{\tau} dt_1 \int_0^{\tau} dt_2 \langle \delta %j(L,t_1) \delta j(L,t_2)\rangle \\
%&=\int d\omega\, S(L,\omega) \cdot  \frac{1-\cos(\omega \tau)}
%{\pi \omega^2 \tau }.
%\label{Variance of charge 1}
%\end{align}
In the long time limit the function $[1-\cos(\omega \tau)]/\pi \omega^2 \tau $ becomes a Dirac $\delta$-function, and thus we obtain
\begin{equation}
\label{Variance of charge 2}
S(L,0)=\lim_{\tau \to \infty} \frac{\langle [\delta Q(L,\tau)]^2\rangle}{\tau}.
\end{equation}

We note, that the right hand side of this equation is conserved, i.e.,  in any finite (and isolated) chiral 1D system it coincides with its value at the injection point, $x=0$. This follows from the charge conservation and from the fact that the fluctuation of charge localized in the finite systems between points $x=0$ and $x=L$, equal to $\delta Q(0,\tau)-\delta Q(L,\tau)$, is limited by its Coulomb energy, while both $\delta Q(L,\tau)$ and $\delta Q(0,\tau)$ scale up as $\sqrt{\tau}$, as follows, e.g., from the Eq.\ (\ref{Variance of charge 2}). Therefore, $S(L,0)=S(0,0)$, and thus  contrary to the expected value (\ref{FDT}) in equilibrium, the zero frequency noise power must take value  $S(L,0)=G_q T_{\rm eff}$, where $T_{\rm eff}$ is so defined effective temperature of the initial non-equilibrium state. However, $T_{\rm eff}\neq T_\infty$ in general (as, e.g., in the case of the current injection via a QPC, see the discussion below), therefore the assumption of full equilibration at any finite distances $L>L_{\rm eq}$ breaks. 

\begin{figure}
\includegraphics[width=\columnwidth]{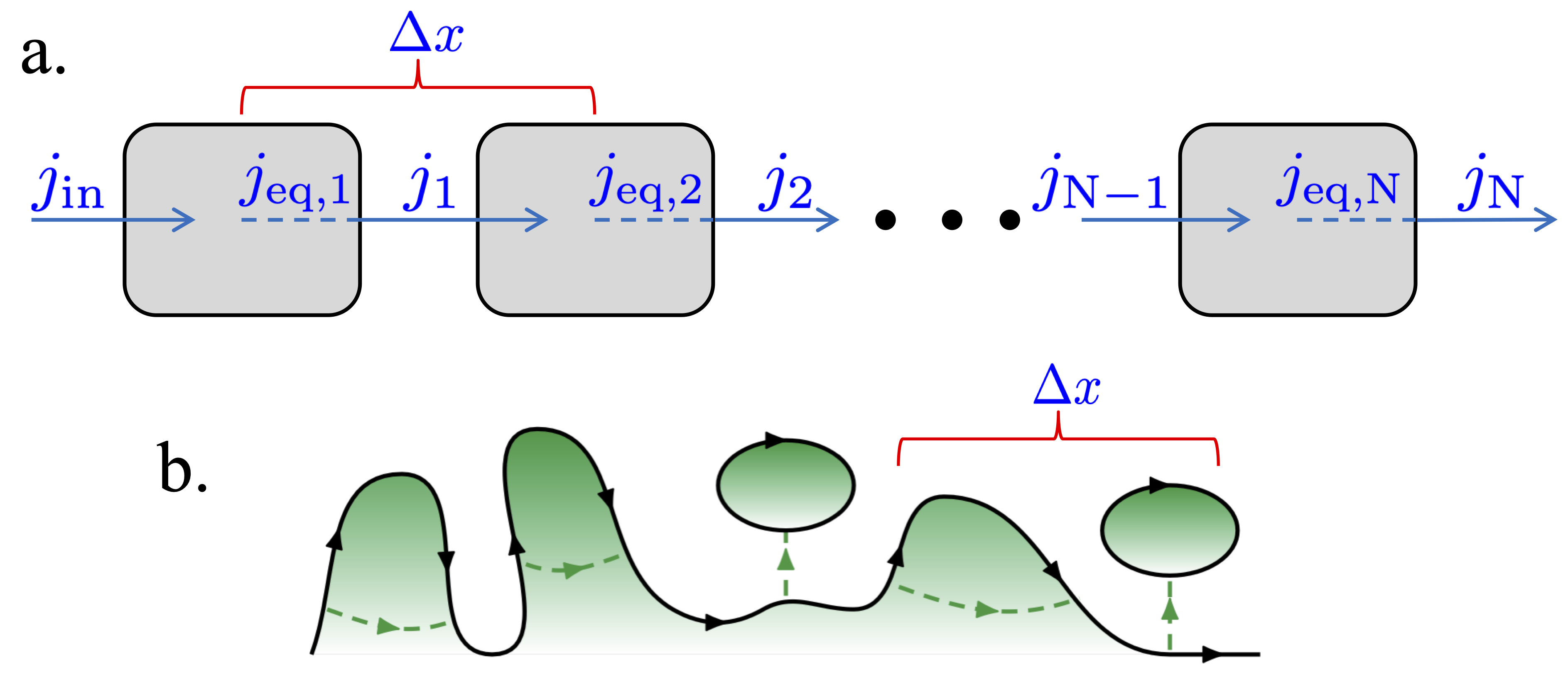}
\caption{\label{fig:two} (a) The transmission line model of the QH edge at filling factor 1 is illustrated. The notations are explained in the main text. (b) A somewhat more realistic picture of a QH edge with disorder is shown. The edge state forms loops and puddles of the characteristic size $\Delta x$ connected by tunneling, shown by dashed lines. }
\end{figure}

From this analysis it follows, that at any finite distance $L$ only low energy excitations are protected by the charge conservation from the equilibration, as the assumption of equilibrium breaks at zero (or sufficiently low) frequencies.
The high energy states most likely equilibrate quite soon after the injection of the current. In this paper we propose an effective way to account such equilibrium states in the overall equilibration process, based on the transmission line model \cite{Stabler}, which is closely related to the Caldeira-Leggett model of dissipative quantum systems \cite{Leggett}. We present the unknown part of the system that is weakly coupled to the edge state and equilibrates it as a set of locally equilibrium reservoirs (see Fig.\ \ref{fig:two}). Coupling to these reservoirs leads also to the dissipation and relaxation of edge states, confirmed experimentally \cite{Bocquillon2013}. We conjecture, that the details of a particular equilibration mechanism do not affect the universality of the equilibration process under the constraint of  the requirement of the charge and energy conservation. Taking advantage of the simplicity of our model, we derive the asymptotic behavior of the electron distribution function at long distances $L$ from the injection point and demonstrate that its deviation from the eventual equilibrium state scales as $1/\sqrt{L}$, i.e., the characteristic equilibration length is infinite. Throughout the paper, we set $|e|=\hbar=k_B=1$. The details of calculations as well as exact numerical evaluations supporting our theory are presented in appendixes.

  %=================================================================================
\section{Transmission line model}
\label{Transmission line model}
Following the  Ref.\  \cite{Stabler}, we describe a QH edge at filling factor 1 with intrinsic dissipation with the help of the transmission line model, see Fig.\ \ref{fig:two}. While propagating along the edge, the state sequentially enters and exist the metallic reservoirs, equally spaced at the distance $\Delta x$ and all having the same charge capacitance $C$ (The effects of randomness of this parameters, discussed in the end of the paper, is rather minor and can be neglected). We assume a long dwell time of electrons in the reservoirs, so that the state is fully equilibrated inside of them. The generally non-equilibrium current $j_{n-1}$ enters the $n$th reservoir, while the current that exits can be written as $j_n=G_qQ_n/C+j_{{\rm eq},n}$, where  $j_{{\rm eq},n}$ is the equilibrium (with the temperature $T_n$) neutral fermionic current inside the reservoir. The assumption of the long dwell time implies also that currents $j_{n-1}$ and $j_{{\rm eq},n}$ are not correlated. The correlation between currents $j_{n-1}$ and $j_n$ arise via the charge continuity equation $dQ_n/dt=j_{n-1}-j_n$, which in frequency representation gives
\begin{equation}
\label{Single reservoir}	
j_n(\omega)=a(\omega)j_{n-1}+b(\omega)j_{{\rm eq},n},
\end{equation} 
where $a(\omega)=1/(1-i\omega \tau_C)$ is the transmission amplitude, $b(\omega)=1-a(\omega)$, and $\tau_C=C/G_q$ is the charge relaxation time of the reservoirs. Applying this equation  $N $ times, we obtain the edge state current $j_N$ at the distance $L=\Delta x N$ from the injection point (see Fig.\ \ref{fig:two}a):
\begin{equation}
\label{N reservoirs}
j_N(\omega)=a^N(\omega) j_{\rm in}(\omega)+b(\omega)\sum^N_{n=1} a^{N-n}(\omega) j_{{\rm eq},n}(\omega).
\end{equation}
where $j_{\rm in }$ is the incident current. 

In order to find the spectrum of the collective charge excitation in the transmission line, we investigate its response to the initial injected current. This (statistically average) response is described by the Eq.~(\ref{N reservoirs}), where fluctuations, and thus the second term on the right hand side, are neglected \footnote{The small excitation $\Delta j_N$ has to be considered as statistically average response to the initial perturbation, while currents  $j_{{\rm eq},n}$ are fluctuations with zero average (the Langevin sources). They do not contribute to the average fields.}. We are interested in the long wave-length excitations, i.e., in the limit of small $\omega\tau_C$, therefore $a(\omega)$ is close to $1$, and thus we can expand the logarithm to second order to also account for the dissipation effects:  $\ln a=-\ln(1-i\omega\tau_C)=i\omega\tau_C-(\omega\tau_C)^2/2$. As follows from the Eq.~(\ref{N reservoirs}), the contributions of the reservoirs to $\ln j_N$ are additive, so that one contribution can be written as   $\ln a =ik \Delta x$, where $\Delta x$ is the distance between the reservoirs and $k$ is the wave vector. By solving the resulting equation  $\omega\tau_C+i(\omega\tau_C)^2/2=k\Delta x$ perturbatively with respect to $\omega$ at small $k$, one arrives at the expression for the spectrum of the collective mode:
\begin{equation}
\label{Spectrum}
\omega(k)=k\Delta x /\tau_C-i(k \Delta x)^2/2\tau_C.
\end{equation}
The first term on the right hand side describes the propagation with the group velocity $\Delta x/\tau_C$, while the second term stands for the dissipation and decay of the excitation  \footnote{Extra phase factor accumulated due to the propagation of an edge state between adjacent reservoirs can be easily taken into account by multiplying $a(\omega)$ with it. It has a minor and irrelevant effect in the context of our analysis.}. 

The dissipation is indeed weak in the limit $k\Delta x\ll 1$ and the relaxation length can be estimated as $L_0=\Delta x/(\omega\tau_C)^2$, where $\omega$ is to be replaced by the characteristic energy scale of the problem. One can immediately see, that $kL_0\gg 1$ in the limit $\omega\tau_C\ll 1$. Formally, we are allowed to take the limit of $\Delta x\to 0$ by keeping the group velocity $\Delta x /\tau_C$ constant, assuming the capacitance $C$ scales linearly with $\Delta x$.  However, the decay rate would vanish in this limit, while experimentally it has been shown to be finite 
\footnote{The estimate $\Delta x\approx 400 nm$ obtained by comparing the prediction of the Eq.\ (\ref{Spectrum})  with the results  of the experiment in Ref.\ \cite{Bocquillon2013}  agrees relatively well with the correlation length of disorder at the edge of QH systems based on GaAs.}. Thus, in the following discussion we assume $\Delta x$ to be finite and take the limit of small temperatures $T_{\rm eff}\tau_C\ll 1$.

%=================================================================================
\section{Spectral density of current noise}
\label{Spectral density of current noise}
Formally, the transmission line model can be viewed as the Langevin equation theory \cite{ArturEquil,IdrisovThermalDecay,IdrisovDrag}, where the incident current $j_{\rm in}$ and reservoir currents $j_{{\rm eq},n}$ are uncorrelated and considered as Langevin sources, or equivalently, as a scattering theory \cite{SukhorukovScattering} following from the Hamiltonian approach \cite{Matveev}. Independently of the approach, the spectral density of the current fluctuations (noise power) defined as $S_n(\omega)=\int dt e^{i\omega t} \langle\delta j_n(t)\delta j_n(0)\rangle$ immediately follows from the Eq.\ (\ref{N reservoirs}):
\begin{equation}
\label{Noise series}
S_N(\omega)={\cal T}^N(\omega)S_{\rm in}(\omega)+{\cal R}(\omega)\sum_{n=1}^N{\cal T}^{N-n}(\omega)S_{{\rm eq},n}(\omega),
\end{equation}
where $S_{\rm in}(\omega)$ is the noise power of the incident current, $S_{{\rm eq},n}(\omega)=G_q\omega/(1-e^{-\omega/T_n})$ is noise power of the locally equilibrium neutral currents of the reservoirs, and we introduced the transmission ${\cal T}(\omega)=|a(\omega)|^2=1/[1+(\omega\tau_C)^2]$ and reflection ${\cal R}(\omega)=1-{\cal T}(\omega)$ coefficients for currents.

 We are interested in the behavior of noise at long distances $N\gg 1$, and in the limit of small $\omega\tau_C\sim T_{\rm eff}\tau_C\ll 1$. Therefore, one can approximate ${\cal T}^N$ with the exponential function:
\begin{equation}
\label{exponential}
{\cal T}^N(\omega)=e^{-N(\omega\tau_C)^2},\quad N\gg 1. 
\end{equation}
In the limit $N\gg  N_0=1/(\tau_CT_{\rm eff})^2\gg 1$ this exponential function cuts off the narrow frequency interval around zero frequency and thus sets a new energy scale
$\Omega_N=1/(\sqrt{N}\tau_C)\ll T_{\rm eff}$. The new length scale $ L_0=\Delta x  N_0$ is the characteristic length of the partial equilibration, as we show next. Before this distances the weak coupling to reservoirs can be treated perturbatively, and in general the result depends on details of the initial state in a non-universal way. Therefore, we focus on the limit of $L\gg  L_0$, i.e., $N\gg N_0$ and resum the series in Eq.\ (\ref{Noise series}). 

\begin{figure}
\includegraphics[width=\columnwidth]{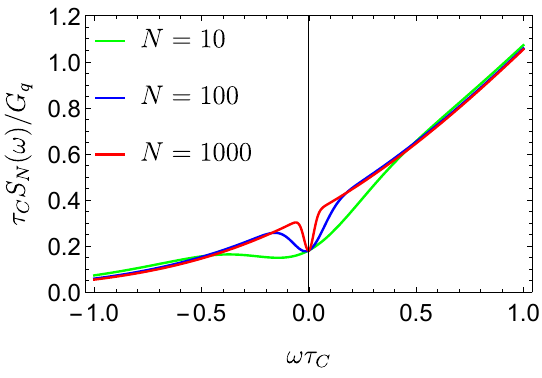}
\caption{\label{Noise power fig} The dimensionless noise power given in  the Eq.\ (\ref{Noise power}) is plotted versus the dimensionless frequency $\omega \tau_C$ for different values of $N$. We set the transparency of the current injecting QPC to $D=0.1$ and the dimensionless bias to $\tau_C \Delta \mu =2$., while $T_N$ is taken from the Eq.\ (\ref{Temperature}) for its asymptotic form.}
\end{figure}

First we note, that at frequencies $\omega\gg \Omega _N$ the first term in Eq.\ (\ref{Noise series}) is exponentially small.  The sum in the second term rapidly converges at the upper limit, assuming that $S_{{\rm eq},n}$ slowly converges towards the limit $S_{{\rm eq},\infty}$. Below we will show that this is indeed the case. Thus we can replace $S_{{\rm eq}, n}$ with its value at $n=N$ and resum the series, which gives $S_N(\omega)=S_{{\rm eq}, N}(\omega)$ for $\omega\gg \Omega _N$. At frequencies $\omega\sim \Omega _N\ll T_{\rm eff}$ the situation is more intricate. The first term in Eq.\ (\ref{Noise series}) takes a simple form $T_{\rm eff}G_q{\cal T}^N$. The sum in the second term can be presented as $G_qT_N(1-{\cal T}^N)+G_q{\cal R}\sum_{n=1}^N{\cal T}^{N-n}(T_n-T_N)$  after resuming the geometrical series. However, the second contribution in this term is sub-leading, if $S_{{\rm eq},n}$ slowly converges towards the limit $S_{{\rm eq},\infty}$, because the factor ${\cal R}$ scales as $1/N$ at $\omega\sim\Omega_N$. Finally, the noise power takes the form
\begin{equation}
\label{Noise power}
S_N(\omega)=G_q(T_{\rm eff}-T_N){\cal T}^N(\omega)+S_{{\rm eq},N}(\omega) 
\end{equation}
where we have omitted sub-leading terms. Thus, the noise power acquires  a sharp peak or deep (depending on the sign of $T_{\rm eff}-T_N$) around zero frequencies that prevents the full relaxation of initial state to the equilibrium at finite distances. In particular, $S_N(0)=S_{\rm in}(0)$, which guarantees the charge conservation. This behavior of the noise power is shown in Fig.\ \ref{Noise power fig}. One can see, that the width of the deep at zero frequencies is slowly decreasing with the distance from the injection point.

%===============================================================================
\section{Energy balance}
\label{Energy balance sec}
The temperature $T_N$ of the $N$th reservoir in Eq.\ (\ref{Noise power}) can be found from the  energy balance equation $\Delta J_{N+1}=\Delta J_{N}$,  where the energy current of a single channel, $\Delta J_{N}=(1/2G_q)\langle j_N^2(t)\rangle=(1/4\pi G_q)\int d\omega \Delta S_N$, is regularized  by subtracting the eventual equilibrium contribution: $\Delta S_{N}\equiv S_{N}-S_{{\rm eq},\infty}$. By substituting $S_N$ from Eq.\ (\ref{Noise power}) in the energy balance equation, one obtains 
\begin{equation}
\label{Energy balance}
\partial_N\int d\omega S_{{\rm eq},N}(\omega)=G_q(T_{\rm eff}-T_{\infty})\int d\omega {\cal R}{\cal T}^N,
\end{equation}
where we neglected sub-leading terms. Calculating the frequency integrals in this equation and integrating over $N$, we arrive at the following result:
\begin{equation}
\label{Temperature}
T_N-T_{\infty}=-\frac{3}{2\pi^{3/2}\sqrt{N}}\frac{T_{\rm eff}-T_{\infty}}{\tau_C T_\infty}.
\end{equation}
Therefore, indeed $T_N$ slowly converges to its limit $T_\infty$. Note also the presence of the small parameter $\tau_CT_\infty$ in the denominator of this expression. It agrees with the earlier made remark that this asymptotic behavior sets at $N\sim N_0=1/(\tau_CT_{\rm eff})^2\gg 1$, i.e., at distances $L\sim L_0=\Delta x/(\tau_CT_{\rm eff})^2$. This characteristic length scale agrees also with the relaxation length that follows directly from the spectrum of the collective charge mode (\ref{Spectrum}).

Finally,  the asymptotic value of the  temperature $T_\infty$ can be found by comparing the heat flux quantum carried by the equilibrium edge state to the injected heat:
\begin{equation}
\label{Injected heat}
J_q=\frac{\pi T_\infty^2}{12}=(1/4\pi G_q)\int d\omega (S_{\rm in}-S_q),
\end{equation}
where we have subtracted the vacuum contribution $S_q=G_q\omega\theta(\omega)$. Two temperatures, $T_{\rm eff}$ and $T_\infty$, are generally different parameters of the noise power of the incident non-equilibrium current $j_{\rm in}$. However, often they are related to each other via the common energy scale. For instance, if the current source is created by injecting electrons into the edge state via a QPC at zero bath temperature, the noise power reads: $S_{\rm in}(\omega)=S_{\rm q}(\omega)+RD\left[\sum_{\pm}S_{\rm q}(\omega \pm \Delta\mu)-2S_{\rm q}(\omega)\right]$, where $D=1-R$ is the transmission probability of the QPC, and $\Delta\mu$ is voltage bias \cite{Blanter}. Thus, $S_{\rm in}(0)=G_qRD\Delta\mu$, and $T_{\rm eff}=RD\Delta\mu$. The integral on the right hand side of the Eq.\ (\ref{Injected heat}) takes the value $(1/4\pi)RD\Delta\mu^2$, so that $T_\infty=\sqrt{(3/\pi^2)RD}\Delta\mu$. Therefore, in this particular case two temperatures differ only by a numerical factor, and $T_\infty>T_{\rm  eff}$, so that $T_N-T_\infty>0$.

%===============================================================================
\section{Electron distribution function}
\label{Electron distribution function}
According to the effective field theory of QH edge states \cite{Wen} and the bosonization technique \cite{Giamarchi} the electron distribution function is given by the expression 
\begin{equation}
\label{Distribution function}	
f_N(\varepsilon) \propto \int dt e^{-i\varepsilon t}\langle e^{-i\phi_N(t)}e^{i\phi_N(0)}\rangle,
\end{equation}
where $e^{i\phi_N}$ represents an electron operator in the channel exiting the $N$th reservoir, and the phase operator is related to the current as $j_N=-(1/2\pi)\partial_t \phi_N$. Assuming the current noise is Gaussian (the non-Gaussian effects are sub-leading in $1/N$ \cite{IvanRelaxation}), the average in this equation reads: 
\begin{multline} 
\label{Phase correlation}	
\ln \langle e^{-i\phi_N(t)}e^{i\phi_N(0)}\rangle=2\pi i \langle j_{\rm in}\rangle t\\ +\langle\phi_N(t)\phi_N(0)-\phi_N^2(t)/2-\phi_N^2(0)/2\rangle.
\end{multline} 
By substituting the Eq.\ (\ref{Noise power}) to this expression and then to Eq.\ (\ref{Distribution function}), we observe that the leading order contribution to the noise power $S_{{\rm eq},N}$ simply returns the equilibrium fermionic distribution function $f_{{\rm eq},N}=1/[1+e^{(\varepsilon-\mu)/T_N}]$, where  $\mu=2\pi  \langle j_{\rm in}\rangle+\varepsilon_F$. By expanding this distribution function in small $T_N-T_{\infty}$ we find the leading order correction  $(T_N-T_{\infty})\partial_{T_\infty}f_{{\rm eq},\infty}$  to the eventual equilibrium distribution $f_{{\rm eq},\infty}$. It scales downs as $1/\sqrt{N}$ according to the result (\ref{Temperature}).

Another leading order correction arises from the first term in Eq.\ (\ref{Noise power}). We note that it is not small only at  frequencies or the order of  $\Omega_N\ll T_{\rm eff}$, while the main contribution in the integral in Eq.\ (\ref{Distribution function}) comes from the leading term $S_{{\rm eq},\infty}$ that sets the time scale $t\sim 1/T_{\rm eff}\ll 1/\Omega_N$. At this time scale the first term in the Eq.\ (\ref{Noise power}) can be considered as the delta function of frequency, which in the time domain gives the correction $G_q(T_{\rm eff}-T_\infty)/2\sqrt{\pi N}\tau_C$  (replacing $T_N\to T_\infty$) to instant fluctuations of the current $\langle\delta j^2(0)\rangle$. Thus, the  phase correlation function on the right hand side of the Eq.\ (\ref{Phase correlation}) acquires the correction $-2\pi^2t^2\langle\delta j^2(0)\rangle$. After expanding the average in the Eq.\ (\ref{Distribution function}) with respect to this correction and evaluating the time convolution with the leading order equilibrium correlation function one obtains the correction $(\sqrt{\pi}/2\sqrt{N}\tau_C)(T_{\rm eff}-T_\infty)\partial^2_{\epsilon}f_{{\rm eq},\infty}$ to the equilibrium distribution $f_{{\rm eq},\infty}$, where we used $G_q=1/2\pi$.  Putting these two corrections together, we obtain our main result:
\begin{equation}
\begin{aligned}  
\label{Final result}	
& f_N-f_{{\rm eq},\infty} =\frac{\sqrt{\pi}}{2}\frac{T_{\rm eff}-T_\infty}{\sqrt{N}\tau_CT_\infty^2} \cdot g(z),\\
& g(z)=\partial^2_zf_F(z)+\frac{3z}{\pi^2}\partial_zf_F(z),\quad z=\frac{\varepsilon-\mu}{T_\infty},
\end{aligned}
\end{equation}
where, we recall,  $f_F(z)=1/(1+e^z)$ and $\mu=2\pi \langle j_{\rm in}\rangle+\varepsilon_F$. 

\begin{figure}
\includegraphics[width=\columnwidth]{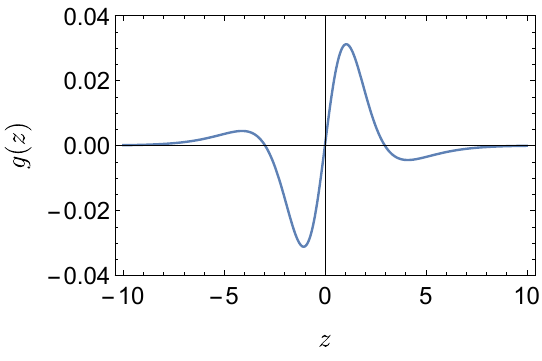}
\caption{\label{Result} The dimensionless function $g(z)$ defined in  the Eq.\ (\ref{Final result}) is plotted versus the normalized energy $z=(\varepsilon-\mu)/T_\infty$.}
\end{figure}

We conclude that the correction to the final equilibrium distribution function scales down slowly as $1/\sqrt{N}$ and is universally expressed in terms of the ``deformations'' of the equilibrium distribution function and of the parameters $T_{\rm eff}$ and $T_\infty$ of the initial state. Thus, the characteristic equilibration length formally is absent, and this fact  resolves the paradox, formulated in the introduction. Nevertheless, the correction (\ref{Final result}) is rather small. Indeed, the overall prefactor in Eq.\ (\ref{Final result}) can be conveniently estimated as $\sqrt{L_0/L}\times(T_{\rm eff}-T_\infty)/T_{\rm eff}$, where $L_0$ is the introduced earlier relaxation length of the collective charge excitation. We recall, that only at distances $L\gg L_0$ the correction to the final distribution function acquires its universal form (\ref{Final result}). Assuming $T_{\rm eff}-T_\infty\sim T_{\rm eff}$ and noticing the maximum value of the function $|g(z)|$ shown in Fig.\ (\ref{Result}), we find that the universal power-law correction does not exceed few percents. Thus, the equilibration mostly takes place at distances $L\sim L_0$. However, this process is not universal as it strongly depends on details of the initial state right after the injection point.

%===============================================================================
\section{Discussion}
\label{Discussion}
It remains to discuss several additional aspects of our theory. First of all, details of the particular kind of coupling of the edge state to reservoirs assumed in our model is likely not important. For instance, one can replace it with just a small capacitive coupling of the channel to  otherwise isolated reservoirs. Then the transmission probability of the current generally behaves as ${\cal T}=1-\alpha\omega^2+\ldots $, where the constant $\alpha$ depends on the strength of  capacitive coupling and is small. Thus in the limit of weak  coupling ${\cal T} $ takes exactly same form as the one used in the paper, and in particular, the limit (\ref{exponential}) still holds. 

Next, if the disorder at the edge of a QH system is responsible for the relaxation process, then it is natural to assume that capacitances of the reservoirs or coupling constants fluctuate from one reservoir to another. 
Assuming that the parameter $\tau_{C,n}^2$ is random function with the average value $\tau_{C}^2$ and variance $\tau_C^4$, in Eq.\ (\ref{exponential}) ${\cal T}^N$ has to be replaced with ${\cal P}_N=\prod_n {\cal T}_n=\exp[-\omega^2\sum_n\tau_{C,n}^2]$. We first account for the self-averaging effect: $\langle {\cal P}_{N}\rangle=\exp[-N(\omega\tau_C)^2+(1/2)N(\omega\tau_C)^4]$, where we average over eventual Gaussian fluctuations of $\sum_n\tau_{C,n}^2$. Thus, we see that at relevant frequencies where $N(\omega\tau_C)^2\sim 1$, the second term in the exponent scales as $1/N$ and gives a sub-leading contribution, that can be neglected. The variance of ${\cal P}_N$ can be easily calculated: $\langle \delta{\cal P}^2_N\rangle=2N(\omega\tau_C)^4\langle {\cal P}_{N} \rangle^2\sim1/N$. Therefore, the typical fluctuation correction to  the initial condition contribution to $S_N$ in Eq.\ (\ref{Noise power}) at relevant frequencies $\omega\sim\Omega_N$ scales as $1/\sqrt{N}$ and can be neglected.

We further note that our results can be  easily generalized to arbitrary integer filling factors, including an experimentally important case of $\nu = 2$. For example, in the latter case, the transmission line model applies separately to the
dipole and charged mode, if tunneling between QH edge channels is negligible.  However, if the injection and detection takes place in the same QH channel, one needs to account for the fact that for strong Coulomb interactions such channel  has equal coupling to both modes. In this case the parameter $1/\tau_C$ in the result   (\ref{Final result}) splits in the sum of two contributions from the two modes divided by $4$.

Finally, we would like to mention that our theory in its present form cannot be directly  apply to non-chiral edge states, and in particular, to fractional QH systems with reconstructed edge states. In these systems the back scattering processes lead to the diffusion of 
charge or heat (or both), so that the initial current state decays much faster than in chiral edge states. Moreover, the paradox formulated in the introduction, being  based on the charge conservation, does not apply in a non-chiral case.  However, clearly the reduced dimensionality of non-chiral edge states should strongly affect the equilibration process, and the transmission line model, after some modifications~\cite{SukhorukovLast}, can still be used to study these systems.

%========================================================  
\section{Conclusion}
\label{Conclusion}
To summarize, we addressed an experimentally relevant problem of the equilibration of chiral QH edge state. We started by presenting a simple physical argument that shows  that the assumption of characteristic time scale for the equilibration breaks down, because it contradicts the charge conservation requirement. We used a transmission line model that effectively accounts for the dissipation and equilibration to find the spectral density of current noise. Then we applied the bosonization technique to evaluate corrections to the locally equilibrium electron distribution function. We showed that the correction scales down slowly as $\sqrt{L_0/L}$ as a function of the distance $L$, where $L_0$ is the relaxation length of the charged mode.

\section*{Acknowledgments}     
We are grateful to Florian St\"abler for fruitful discussions. 
EGI acknowledges financial support from the United Arab Emirates University under the Startup Grant/G00004974. EVS acknowledges the financial support from the Swiss National Science Foundation. 

\appendix

\begin{widetext}

\section{Using energy balance equations for finding temperatures of metallic reservoirs and their exact numerical evaluation}
\label{Sec: III}
 Taking into account the Joule's law, we write the correction to the  energy flux in the channel exiting form the $N$th reservoir, 
 \begin{equation}
\label{Supp: Energy balance equation through noise 1}
\Delta J_N=(1/2G_q)\langle j^2_N(t)\rangle=(1/4\pi G_q) \int d\omega \Delta S_N(\omega),\quad \Delta S_N(\omega)=S_N(\omega)-S_{{\rm eq,}\infty}(\omega),
\end{equation}
by subtracting the eventual equilibrium spectral density in the integrand  to regularize the integral. The energy flux is conserved in the stationary case considered here,
$\Delta J_{N+1}=\Delta J_N$, therefore one can write: $\int d\omega \left(S_{N+1}(\omega)-S_{N}(\omega)\right)=0$. By substituting here the Eq.\ (\ref{Noise power}) from the main text of the paper, we arrive at the following expression: 
\begin{equation}
\label{Supp: Energy balance equation through noise 2}
\int d\omega \left(S_{{\rm eq},N+1}(\omega)-S_{{\rm eq},N}(\omega)\right)=\int d\omega G_q \left(T_{\rm eff}-T_N\right) \cdot \mathcal{R}(\omega)\mathcal{T}^N(\omega)+\int d\omega G_q \left(T_{N+1}-T_N\right) \cdot \mathcal{T}^{N+1}(\omega).
\end{equation}
In the limit $N \gg 1$ one can neglect the last term on the right hand side, since it is a sub-leading contribution in the vicinity of $T_N \to T_{\infty}$. For the same reason $T_N$ can be replaced with $T_\infty$ in the first term. Assuming additionally that in in the limit $N\gg 1$ the temperature $T_N$ is a smooth function of $N$, one can write $S_{{\rm eq},N+1}(\omega)-S_{{\rm eq},N}(\omega)=\partial S_N(\omega)/\partial N$, and we arrive at the Eq.~(\ref{Energy balance}) of the  main text. Therefore,  Eq.~(\ref{Energy balance}) is used to find the temperature in the Eq.~(\ref{Temperature}). 

In what follows, we present numerical calculation of the temperature $T_N$ that support our analytical solution given by  the Eq.~(\ref{Temperature}) in the main text. For doing so, we use energy balance equation in the form $\Delta J_N=\Delta J_{\rm in}$ where $\Delta J_N$ is defined earlier in Eq.\ (\ref{Supp: Energy balance equation through noise 1}), and similarly,  $\Delta J_{\rm in}=(1/4\pi G_q)\int d\omega \Delta S_{\rm in}(\omega)$. Using the exact relation given by the  Eq.~(\ref{Noise series}) of the  main text we can write
\begin{equation}
\label{Supp: Incoming heat equals to outgoing heat}	
\int d\omega \mathcal{T}^N(\omega)\Delta S_{\rm in}(\omega)+\sum_{n=1}^N \int d\omega \mathcal{R}(\omega)\mathcal{T}^{N-n}(\omega) \Delta S_{{\rm eq},n}(\omega)=0,
\end{equation}
where $ \Delta S_{{\rm eq},n}=S_{{\rm eq},n}(\omega)-S_{{\rm eq},\infty}(\omega)$.
It turns out to be more convenient to redefine frequency integrals in the above equation by subtracting from each spectral density the vacuum part  $S_q(\omega)=G_q \omega \theta(\omega)$ and introducing the overline symbol instead of $\Delta$: 
\begin{equation}
\label{Supp: Chain of equations for determining the temperature1}
 \sum_{n=1}^N \int d\omega \mathcal{R}(\omega)\mathcal{T}^{N-n}(\omega) [\overline S_{{\rm eq},n}(\omega)-\overline S_{{\rm eq},\infty}(\omega)]\\
=-\int d\omega \mathcal{T}^N(\omega)\left[\overline S_{\rm in}(\omega)-\overline S_{{\rm eq},\infty}(\omega)\right],
\end{equation}
where, for instance, $\overline S_{\rm in}(\omega)=S_{\rm in}(\omega)-S_{q}(\omega)$. 
For a numerical procedure, we need to specify the function $S_{\rm in}(\omega)$, i.e., the source of the current noise. We choose to consider the injection via a QPC with the the transparency $D$, biased by the electrochemical potential $\Delta\mu$ (see the discussion in the main text of the paper). In this case $\overline S_{\rm in}(\omega)=D(1-D)[\sum_\pm S_{q}(\omega\pm\Delta\mu)-2S_{q}(\omega)]$ with $T_{\rm eff}=D(1-D) \Delta \mu$ and $T_\infty=\sqrt{(3/\pi^2)RD}\Delta\mu$.

Next, introducing the dimensionless variable $y=\omega \tau_C$, we obtain the chain of equations for the determination of the temperatures $T_N$ of Ohmic contacts:
\begin{equation}
\label{Supp: Chain of equations for determining the temperature2}	
\begin{split}
& F_{{\rm eq}, \infty}(N,\tau_C T_{\infty})-\Phi_{\rm in}(N,\Delta \mu \tau_C, D)=\\
& =\sum_{n=1}^N \left[F_{{\rm eq}, n}(N-n,\tau_C T_{n})-F_{{\rm eq}, n}(N-n+1,\tau_C T_{n})\right]-\sum_{n=1}^N\left[F_{{\rm eq}, n}(N-n,\tau_C T_{\infty})-F_{{\rm eq}, n}(N-n+1,\tau_C T_{\infty})\right], \\
\end{split}
\end{equation}		
where the dimensionless function are given by 
\begin{equation}
\label{Supp: Dimensionless functions}
\begin{split}
& \Phi_{\rm in}(N,\Delta \mu \tau_C, D)=\int \frac{(1-D)Ddy}{(1+y^2)^N}\left[S_0(y+\Delta \mu \tau_C)+S_0(y-\Delta \mu \tau_C)-2S_0(y)\right], \quad S_0(y)=y\theta(y),\\
& F_{{\rm eq}, \infty}(N,\tau_C T_{\infty})=\int_0^{\infty} \frac{2ydy}{(1+y^2)^{N}(1-\exp(y/\tau_C T_{\infty}))}, \\
& F_{{\rm eq}, n}(N,\tau_C T_n)=\int_0^{\infty} \frac{2ydy}{(1+y^2)^{N}(1-\exp(y/\tau_C T_n))}.
\end{split}
\end{equation}
To arrive at  Eqs.~(\ref{Supp: Chain of equations for determining the temperature2}) we have not applied any approximations, so they can be used  for the exact numerical evaluation of the temperatures $T_N$. For numerical calculations we iteratively apply brute-force quadrature integration and Powell hybrid method from SciPy package~\cite{Numerics}. The result is shown in Fig.~\ref{fig:One}, where they are compared to the  asymptotic behavior in Eq.~(\ref{Temperature}) of the  main text found analytically. They demonstrate the essentially perfect match of the numerical and analytical results for asymptotically large values of $N$. However, the convergence is rather slow, which can be explained by the existence of subleading power law contributions that scale down as $1/N$ and faster.

\begin{figure}
\includegraphics[width=\textwidth]{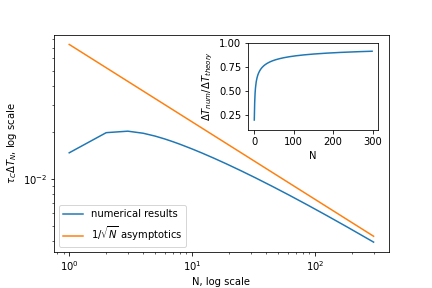}
\caption{\label{fig:One} The comparison of numerical and analytical results for dimensionless temperature, $\tau_C \Delta T_N=\tau_C(T_N-T_{\infty})$ are shown. The results are presented in a log scale to demonstrate the $1/\sqrt{N}$ behavior in the limit $N \gg 1$, predicted by  the Eq.~(\ref{Temperature}) in the main text. We set the transparency of the current injecting QPC to $D=0.2$ and  the dimensionless bias to $\tau_C \Delta \mu=2$, so that the dimensionless effective temperature of initial noise acquires the value $\tau_C T_{\rm eff}=\tau_C D(1-D) \Delta \mu=0.32$.  The inset shows the ratio between numerical and theoretical results.}
\end{figure}

\section{Non-equilibrium distribution function}
\label{Sec: IV}
In this section we derive the Eq.~(\ref{Final result}) from the main text. To do so, we express bosonic fields in currents $\phi_N(t)=-2\pi \int_{-\infty}^t dt_1 j_N(t_1)e^{\eta t_1}$, where $\eta \to +0$. Then, the right hand side of the  Eq.~(\ref{Phase correlation}) in the main text reads 
\begin{equation}
\label{Supp: Average of bosonic fields is presented through currents}
\begin{split}
&\langle \phi_N(t) \phi_N(0)- \phi^2_N(t)/2- \phi^2_N(0)/2\rangle\\
&\qquad\qquad=(2\pi)^2 \int_{-\infty}^t dt_1 \int^0_{-\infty} dt_2 e^{\eta(t_1+t_2)} \left[\langle \delta j_N(t_1) \delta j_N(t_2)\rangle-\langle \delta j^2_N(t_1)\rangle/2-\langle \delta j^2_N(t_2)\rangle/2\right].
\end{split}
\end{equation}
Next, using the definition of the noise power of currents, $S_N(\omega)=\int dt e^{i\omega t} \langle \delta j_N(t) \delta j_N(0)\rangle$, the above equation can be written as 
\begin{equation}
\label{Supp: Average of bosonic fields is presented through noise power}	
\langle \phi_N(t) \phi_N(0)- \phi^2_N(t)/2- \phi^2_N(0)/2\rangle = 2\pi \int\limits_{-\infty}^\infty d\omega \frac{e^{-i\omega t}-1}{\omega^2+\eta^2} S_N(\omega).
\end{equation}
Using this result in the  Eq.~(\ref{Distribution function}), we arrive at the following expression for the electron distribution function
\begin{equation}
\label{Supp: Distribution function 1}
f_N(\epsilon) \propto \int \limits_{-\infty}^\infty dt \exp \left[-i(\epsilon-2\pi \langle j_{\rm in} \rangle)t+ 2\pi \int_{-\infty}^\infty d\omega \frac{e^{-i\omega t}-1}{\omega^2+\eta^2} S_N(\omega)\right],
\end{equation}
where the prefactor coefficient can be fixed by comparing to the equilibrium distribution.

In the next step, we substitute the Eq.~(\ref{Noise power}) for the  noise power from the main text into the above equation and split the integrand in the product of different contributions:
\begin{equation}
\label{Supp: Distribution function 2}
\begin{split}
  f_N(\epsilon) \propto \int\limits_{-\infty}^\infty dt \exp\left[-i(\epsilon-2\pi \langle j_{\rm in}\rangle)t\right] & {\;}\exp\left[2\pi \int_{-\infty}^\infty d\omega \frac{e^{-i\omega t}-1}{\omega^2+\eta^2} S_{{\rm eq},N}(\omega)\right] \\
&\times\exp \left[2\pi G_q(T_{\rm eff}-T_N) \int_{-\infty}^\infty d\omega \frac{e^{-i\omega t}-1}{\omega^2+\eta^2}\mathcal{T}^N(\omega)\right].
\end{split}
\end{equation}
The second exponential function in  the integrand is nothing but the equilibrium free-fermionic correlation function with the temperature $T_N$ found earlier,
\begin{equation}
\label{Supp:Free-fermion correlation function}
\exp\left[2\pi \int_{-\infty}^\infty d\omega \frac{e^{-i\omega t}-1}{\omega^2+\eta^2} S_{{\rm eq},N}(\omega)\right]=\frac{-iT_N}{2\sinh(\pi T_N(t-i\eta))}\equiv K_{{\rm F}, N}(t).
\end{equation} 
Taken alone (together with the first exponential function in the integrand ) the function $K_{{\rm F}, N}(t)$  gives the dominant contribution to the integral (\ref{Supp: Distribution function 2}), namely, $f_{{\rm eq},N}(\epsilon)=(1+e^{(\epsilon-\mu)/T_N})^{-1}$, where $\mu=2\pi \langle j_{\rm in}\rangle$,
and thus it fixes the characteristic times scale $1/T_N$ in the time integral. 

On the  corresponding frequency scale $\omega\sim T_N$ the function 
$\mathcal{T}^N(\omega)$ in the exponent of the third exponential function in (\ref{Supp: Distribution function 2}) can be replaced by delta function $\sqrt{\pi} \delta(\omega)/\sqrt{N} \tau_C$. Thus we obtain:
\begin{equation}
\label{1}
  \exp \left[2\pi G_q(T_{\rm eff}-T_N) \int_{-\infty}^\infty d\omega \frac{e^{-i\omega t}-1}{\omega^2+\eta^2}\mathcal{T}^N(\omega)\right] 
  =\exp\left[-2\pi G_q (T_{\rm eff}-T_N)\frac{\sqrt{\pi} t^2}{2\sqrt{N}\tau_C}\right].
\end{equation}
Since $t\sim 1/T_N$, the argument of the exponential function on the right hand side of this equation can be estimated as $1/T_N\tau_C\sqrt{N}$ and becomes small in the limit of interest: $N\gg N_0=1/(\tau_CT_{\rm eff})^2$ (see the discussion in the main text). Therefore, we can expand the exponential functions and keep only two terms:
\begin{equation}
\label{2}
  \exp \left[2\pi G_q(T_{\rm eff}-T_N) \int_{-\infty}^\infty d\omega \frac{e^{-i\omega t}-1}{\omega^2+\eta^2}\mathcal{T}^N(\omega)\right] 
  =1-2\pi G_q (T_{\rm eff}-T_N)\frac{\sqrt{\pi} t^2}{2\sqrt{N}\tau_C}.
\end{equation}
The first term on the right hand side gives the dominant contribution $f_{{\rm eq},N}(\epsilon)$ to the distribution function (\ref{Supp: Distribution function 2}), discussed earlier, while the second term gives a small correction to it. The result of the time integration can be presented as: 
\begin{equation}
\label{Supp: Distribution function 5}
f_N(\epsilon)=f_{{\rm eq},N}(\epsilon)+\frac{T_{\rm eff}-T_{N}}{T^2_{N}}\frac{\sqrt{\pi}}{2\sqrt{N}\tau_C} \frac{\partial^2 f_{\rm F}(z)}{\partial z^2}, \qquad z=(\epsilon-\mu)/T_{N}.
\end{equation}
Finally, we note, that the correction on the right hand side scales as $1/\sqrt{N}$, while the correction to the temperature $\Delta T_N=T_N-T_\infty$ also scales as $1/\sqrt{N}$, as has been found in the previous section. Therefore, the first term on the right hand side of the above equation can be expanded to first order in the correction, while in the second term we can simply replace $T_N$ with $T_\infty$. The result reads:
\begin{equation}
\label{Supp: Distribution function 6}
f_N(\epsilon) = f_{{\rm eq},\infty}(\epsilon)+\frac{T_{\rm eff}-T_{\infty}}{T^2_{\infty}}\frac{\sqrt{\pi}}{2\sqrt{N}\tau_C} \left[\frac{3z}{\pi^2}\frac{\partial f_{\rm F}(z)}{\partial z}+\frac{\partial^2 f_{\rm F}(z)}{\partial z^2}\right], \qquad z=(\epsilon-\mu)/T_{\infty}.
\end{equation}
which completes the derivation of the Eq.~(\ref{Final result}) in the  main text.
\end{widetext}
%=========================================================
\bibliography{refs}

\end{document}